\documentclass[prd,aps,twocolumn,a4paper,floatfix,superscriptaddress]{revtex4}

\usepackage{graphicx,psfrag}
\usepackage{mathrsfs}
\usepackage{amsmath,amsfonts,amssymb,amsthm}
\usepackage{url}
\usepackage{color}
\usepackage{float}
\usepackage{ulem}

\begin{document}

\title{Collapse of Nonlinear Gravitational Waves in Moving-Puncture 
Coordinates}

\author{David Hilditch}
\affiliation{Friedrich-Schiller-Universit\"at Jena, 07743 Jena, Germany}

\author{Thomas W. Baumgarte}
\affiliation{Max-Planck-Institute f{\"u}r Astrophysik,
Karl-Schwarzschild-Str.~1, D-85748, Garching bei M{\"u}nchen, Germany}
\affiliation{Department of Physics and Astronomy, Bowdoin College, Brunswick, ME 04011, USA}

\author{Andreas Weyhausen}
\affiliation{Friedrich-Schiller-Universit\"at Jena, 07743 Jena, Germany}

\author{Tim Dietrich}
\affiliation{Friedrich-Schiller-Universit\"at Jena, 07743 Jena, Germany}

\author{Bernd Br\"ugmann}
\affiliation{Friedrich-Schiller-Universit\"at Jena, 07743 Jena, Germany}

\author{Pedro J. Montero}
\affiliation{Max-Planck-Institute f{\"u}r Astrophysik,
Karl-Schwarzschild-Str.~1, D-85748, Garching bei M{\"u}nchen, Germany}

\author{Ewald M\"{u}ller}
\affiliation{Max-Planck-Institute f{\"u}r Astrophysik,
Karl-Schwarzschild-Str.~1, D-85748, Garching bei M{\"u}nchen, Germany}

\begin{abstract}
We study numerical evolutions of nonlinear gravitational waves in 
moving-puncture coordinates. We adopt two different types of initial data -- 
Brill and Teukolsky waves -- and evolve them with two independent codes 
producing consistent results. We find that Brill data fail to produce 
long-term evolutions for common choices of coordinates and parameters, 
unless the initial amplitude is small, while Teukolsky wave initial 
data lead to stable evolutions, at least for amplitudes sufficiently 
far from criticality. The critical amplitude separates initial data 
whose evolutions leave behind flat space from those that lead to a
black hole. For the latter we follow the interaction of the 
wave, the formation of a horizon, and the settling down into a 
time-independent trumpet geometry.   We explore the differences between
Brill and Teukolsky data and show that for less common choices of the parameters -- 
in particular negative amplitudes -- Brill data can be evolved with moving-puncture
coordinates, and behave similarly to Teukolsky waves.  
\end{abstract}

\pacs{
  95.30.Sf,   
  04.25.D-   
}

\maketitle

\section{Introduction}
\label{section:Introduction}

The first successful simulations of binary black hole merger 
and coalescence \cite{Pre05,CamLouMar05,BakCenCho06} marked a 
massive break-through in the field of numerical relativity.  
Since then, numerous simulations of black hole binaries have 
produced important results, including predictions of the 
emitted gravitational wave forms for various binary 
configurations. Many of these simulations have been performed 
with some version of the Baumgarte-Shapiro-Shibata-Nakamura (BSSN) 
formulation~\cite{NakOohKoj87,ShiNak95,BauSha98}. Their success
depends crucially on the use of suitable coordinate conditions. 
For simulations of compact objects, the 1+log slicing 
condition~\cite{BonMasSei94} together with the Gamma-driver 
shift condition~\cite{AlcBruDie02,MetBakKop06} have proven 
versatile. The combination is often referred to as 
``moving-puncture" coordinates.

The geometric properties of moving-puncture coordinates have 
been analyzed by~\cite{HanHusPol06,Bro07a,HanHusOMu06,HanHusOhm08,Bru09}. 
These studies showed that dynamical evolutions of a Schwarzschild
black hole result in spatial slices that do not encounter the central
singularity, and instead asymptote to a finite areal radius. In a
Penrose diagram, these slices connect spatial infinity in one universe
with time-like infinity in the other. In an embedding diagram (see,
e.g.~Fig.~2 in~\cite{HanHusOhm08}) the slices resemble a trumpet,
which explains why they are called ``trumpet" slices.

It is not obvious that, in general, moving-puncture coordinates 
work well for regular initial data that collapse to a black hole. 
Following earlier work~\cite{BaiRez06}, two independent 
calculations~\cite{ThiBerHil10,StaBauBro11} considered stellar 
collapse and found that, in the cases considered, moving-puncture 
coordinates can indeed lead to stable evolution, with the newly-formed 
black hole expressed in a trumpet geometry. The primary motivation 
for this paper is to answer the question: {\it in what scenario, 
if any, can the collapse of gravitational waves to a black hole be 
followed in moving-puncture coordinates?} To do this we consider 
axisymmetric vacuum data of two types, namely Brill and Teukolsky 
waves \cite{Bri59,Teu82}.

A secondary motivation comes from the context of critical collapse 
(see \cite{Gun00,Gun02} for reviews.) Critical collapse in gravitational 
systems was first discovered by Choptuik~\cite{Cho93}, who considered 
scalar fields in spherical symmetry. Parametrizing the strength of the 
initial data with some parameter, say~$A$, it is found that for 
sufficiently small~$A$ the fields ultimately propagate to infinity and 
leave behind flat space, possibly after interacting in a nonlinear 
fashion. Above a critical value~$A_\star$ of the amplitude, however, the 
fields collapse and form a black hole. In the vicinity of~$A_\star$ the 
solution displays critical behavior familiar from other fields of 
physics.

\begin{table*}[t]
  \centering
  \begin{ruledtabular}
  \begin{tabular}{l|l|l|l|l|l}
    \hline
Authors & Year & Data type  & Slicing and gauge & references  & comments \\
\hline
Eppley & 1978 & Brill & maximal slicing/quasiisotropic & \cite{Epp79} & small amplitude waves only \\
Abrahams \& Evans & 1992 &Teukolsky & maximal slicing/quasiisotropic & \cite{AbrEva92,AbrEva93} & reported critical behavior \\
Alcubierre {\it et.al.} & 2000 & Brill  & maximal slicing/zero shift & \cite{AlcAllBru99a} \\
Garfinkle \& Duncan & 2001 &Brill & maximal slicing/quasiisotropic & \cite{GarDun00} \\
Santamaria  & 2006 & Brill & multiple choices   & \cite{San06} \\
Rinne & 2008 & Brill & maximal slicing/quasiisotropic & \cite{Rin08} \\
Sorkin & 2011 & Brill & family of gauge source functions & \cite{Sor10} & reported critical behavior \\
\hline
\end{tabular}
  \end{ruledtabular}
\caption{Summary of published results on numerical simulations of nonlinear waves.}
\label{Table:PreviousResults}
\end{table*}

Similar behavior was found in other gravitational systems. Of 
relevance here, Abrahams and Evans~\cite{AbrEva92,AbrEva93} 
reported critical phenomena in the collapse of axisymmetric 
gravitational waves. Unfortunately it has proven difficult 
to reproduce these results. Various authors have studied the 
evolution of gravitational wave initial data, see 
Table~\ref{Table:PreviousResults} for a list of published results, 
but only Sorkin~\cite{Sor10} has been able to identify critical 
behavior. Even his study, which adopted Brill wave initial 
data~\cite{Bri59,Epp77} and used a generalized harmonic 
code~\cite{Fri85,Gar01,Pre05} in axisymmetry~\cite{Sor09}, required 
fine-tuning of free parameters in the gauge source functions 
that specify the coordinates. Sorkin found qualitative 
differences from the earlier work. For example, he reports that, 
at least for part of the parameter space, the waves collapse to 
form a singularity on a ring in the equatorial plane, whereas 
Abrahams and Evans~\cite{AbrEva93} found the singularity to form 
at the center. Sorkin also found a significantly 
larger value for the critical 
amplitude than in earlier studies of the same data.

Following the first question, our final aim is to report that 
under evolution Brill data~\cite{Bri59,Epp77} behave 
differently from Teukolsky data~\cite{Teu82}, at least for 
common choices of the parameters.  These Brill waves 
fail to produce stable, long-term evolutions when evolved 
with moving-puncture coordinates, unless the initial amplitude is 
small, while Teukolsky data lead to stable evolutions, unless the 
initial amplitude is close to criticality. The obvious question 
is: {\it why is it so difficult to evolve Brill wave data?} 
To {\it begin} to address this question we identify qualitative differences 
in the two initial data types, and examine the effect of these differences on the evolution. 
We also note that the original studies of Abrahams and 
Evans~\cite{AbrEva92,AbrEva93} adopted Teukolsky wave initial data. 
Since then, all published work that we are aware of has used Brill 
wave initial data (see Table~\ref{Table:PreviousResults}) -- perhaps 
because the latter can be constructed more easily. This choice may 
have contributed to the difficulty of studying critical phenomena in 
vacuum spacetimes.

The paper is organized as follows. In Sec.~\ref{section:3+1} we 
review relevant results from the~$3+1$ split of spacetime. 
In Sec.~\ref{section:Waves} we discuss the construction and 
evolution of Brill and Teukolsky wave data. In 
Sec.~\ref{section:Character} we discuss some possible 
causes of the differences in behavior between the two types. In 
Sec.~\ref{section:Summary} we summarize. 
Appendix~\ref{section:Numerics} contains a description of the 
codes employed. We adopt geometrized units~$G = c = 1$.

\section{The 3+1 decomposition}
\label{section:3+1}

\paragraph*{Space-time split:} We solve Einstein's field 
equations in vacuum with the help of a~$3+1$ decomposition 
(\cite{ArnDesMis04,Yor79}, see~\cite{Alc08,BauSha10,Gou12} 
for pedagogical introductions.) We write the spacetime metric
in the form,
\begin{align}
{\rm d}s^2 & = g_{ab} {\rm d}x^a {\rm d}x^b \nonumber \\
& = -\alpha^2 {\rm d}t^2 + \gamma_{ij} ({\rm d}x^i 
+ \beta^i{\rm d}t)({\rm d}x^j + \beta^j {\rm d}t),
\end{align}
where~$\alpha$ is the lapse function,~$\beta^i$ the shift vector, 
and~$\gamma_{ij}$ the spatial metric. Here and in the following 
indices~$a, b, \ldots$ run over spacetime indices, while 
indices~$i,j,\ldots$ run over space indices only. Einstein's 
field equations split into two sets, namely the Hamiltonian and
momentum constraints and the evolution equations.

\paragraph*{Constraint and evolution equations:} The Hamiltonian and momentum 
constraints are given by,
\begin{align} 
\label{Ham}
&R + K^2 - K_{ij} K^{ij} = 0,\\
\label{mom}
&D_j K^{ij} - D^i K = 0.
\end{align}
Here~$R$ is the trace of the Ricci tensor $R_{ij}$ associated 
with the spatial metric~$\gamma_{ij}$, $D_i$ is the covariant 
derivative associated with~$\gamma_{ij}$, $K_{ij}$ is the 
extrinsic curvature
\begin{align} \label{ext_curv}
K_{ij} = - \frac{1}{2 \alpha} \partial_t \gamma_{ij} + D_{(i} \beta_{j)},
\end{align}
and~$K = \gamma^{ij} K_{ij}$ its trace.

Eq.~(\ref{ext_curv}) can 
be solved for the time derivative of the spatial metric~$\gamma_{ij}$, 
which provides one of the two evolution equations. A second 
evolution equation results from Einstein's equations and 
determines the time derivative of the extrinsic curvature. 
We construct numerical solutions to the constraint and evolution 
equations, using the two independent codes described in 
Appendix~\ref{section:Numerics}. We employ a conformal 
transformation of the spatial metric,
\begin{align}
\gamma_{ij} = \psi^4 \bar \gamma_{ij},
\end{align}
where~$\psi$ is a conformal factor and $\bar \gamma_{ij}$ a 
conformally related metric.  The Hamiltonian constraint can 
then be written as an elliptic equation for the conformal 
factor,
\begin{align} \label{Ham_conformal}
\bar D^2 \psi = \frac{\psi}{8} \bar R + \frac{\psi^5}{8} 
(K^2 - K_{ij} K^{ij}),
\end{align}
where~$\bar D^2$ and~$\bar R$ are the Laplace operator and 
Ricci scalar associated with~$\bar \gamma_{ij}$. Solving 
the constraints results in data describing the 
gravitational fields at one instant of time. 

\paragraph*{Gauge choice:} The lapse function~$\alpha$ and the 
shift vector~$\beta^i$ encode the coordinate freedom, and can 
be chosen freely.  Most often we use the~$1+\log$ slicing 
condition~\cite{BonMasSei94}, and Gamma-driver 
shift~\cite{AlcBruDie02,MetBakKop06} conditions,
\begin{align} 
\label{1+log}
(\partial_t - \beta^j \partial_j ) \alpha &= - 2 \alpha K,\\
\label{Gamma-driver}
 (\partial_t - \beta^j \partial_j) \beta^ i &= \mu_S \bar\Gamma^i
-\eta\beta^i,
\end{align}
where~$\bar\Gamma^i\equiv \bar \gamma^{jk}\bar\Gamma^i_{jk}$,
are called conformal connection functions. Here~$\bar\Gamma^i_{jk}$ 
are the connection coefficients associated with the conformally 
related metric~$\bar\gamma_{ij}$. At~$t=0$ we initialize the lapse 
and shift with~$\alpha = 1$ and~$\beta^i=0$. Together, the two 
conditions are usually called ``moving-puncture" coordinates. We 
consider alternatives to~\eqref{1+log} and~\eqref{Gamma-driver}; 
in particular we use a ``non-advective" version of the above 
conditions, in which the shift terms on the left-hand sides are 
omitted, and evolutions with zero shift.

\section{Numerical evolutions of Brill and Teukolsky waves}
\label{section:Waves}

\subsection{Brill Waves}
\label{subsection:Brill}

\paragraph*{Initial data:} Brill wave initial data~\cite{Bri59,Epp77} 
can be constructed by writing the spatial metric~$\gamma_{ij}$ in 
the form,
\begin{align} \label{Brill_metric}
{\rm d}l^2 &= \gamma_{ij} {\rm d}x^i {\rm d}x^j 
=\psi^4 \left[ e^{2 q} ({\rm d}\rho^2 + {\rm d}z^2) 
+ \rho^2  {\rm d}\phi^2 \right].
\end{align}
Here~$\rho$, $z$ and~$\phi$ are cylindrical coordinates,~$\psi$ is 
the conformal factor, and~$q = q(r,\theta)$ is an arbitrary 
axisymmetric seed function. Under the further assumption of 
time-symmetry the momentum constraint~\eqref{mom} is solved 
identically, and the Hamiltonian constraint~\eqref{Ham} reduces 
to a linear elliptic equation for~$\psi$,
\begin{align} \label{Brill_psi}
\nabla^2 \psi = - \frac{\psi}{4} \left( \frac{\partial^2 q}{\partial \rho^2} 
+ \frac{\partial^2 q}{\partial z^2} \right),
\end{align}
where~$\nabla^2$ denotes the flat, three-dimensional Laplace operator.
For our applications we choose the seed function~$q$ according to
\begin{align} \label{Brill_seed}
q(\rho,z) = A \left(\frac{\rho}{\sigma}\right)^2 e^{-[(\rho - \rho_0)^2-z^2]/\sigma^2},
\end{align}
where~$A$ is a measure of the resulting wave amplitude,~$\sigma$ 
of the wavelength, and~$\rho_0$ of the center of the initial wave.  
Throughout we set~$\sigma=1$, which determines the units of all 
dimensional results. Given choices for these parameters we solve 
equation~\eqref{Brill_psi} for~$\psi$ and insert the result, together 
with~$q$, into the metric~\eqref{Brill_metric}. We experiment with 
different values of~$A$ and~$\rho_0$ in what follows. For a given 
choice of the parameter~$\rho_0$ in the Brill seed 
function~\eqref{Brill_seed}, the resulting spacetime depends on 
the amplitude~$A$, although some gauge conditions may be unsuitable 
for computing it. In this section we will focus on ``central" waves 
with~$\rho_0 = 0$.

\begin{figure}[t]
\centering
\includegraphics[]{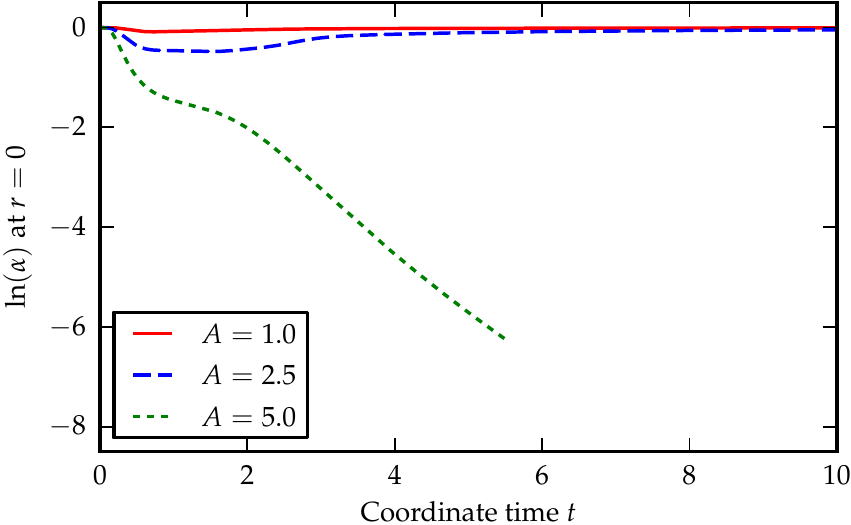}
\caption{The central value of the lapse as a function of time 
for Brill waves with~$\rho_0=0$, evolved with moving-puncture 
gauge conditions. We show results for different values of~$A$.  
For weak-field initial data with~$A=1$, the wave disperses to 
spatial infinity and leaves behind flat space, as expected. For 
larger values of~$A$, however, our simulations develop discontinuities 
in the metric functions, which spoil the further evolution of the 
wave.}\label{Fig:Brill_lapse_at_center}
\end{figure}

\paragraph*{Evolution of~$A=1$ centered Brill data:} 
Following previous attempts in other coordinate systems 
(see Table~\ref{Table:PreviousResults}) we explore the dynamical 
evolution of Brill wave initial data with moving-puncture 
coordinates. For small~$A$, the waves represent a linear 
perturbation of flat space that will propagate to spatial infinity 
and leave behind flat space. As an example let us consider the 
evolution of a centered wave with~$A=1$. For these initial data the 
Kretschmann scalar,
\begin{align}
I&=C_{abcd}C^{abcd}\,,
\end{align}
with the Weyl tensor~$C_{abcd}=R_{abcd}$ in vacuum, takes its maximum value
 of~$\approx 216$ at the origin. We evolve with~$\mu_S=1$ 
and~$\eta=3\approx 1/(10M)$ in the Gamma-driver condition 
(\ref{Gamma-driver}), and find that, at the origin, the lapse 
function decreases initially, but quickly moves back towards unity.  
In Fig.~\ref{Fig:Brill_lapse_at_center} we compare this behavior with 
that for other amplitudes. The initial pulse in the Kretschmann scalar 
disperses away and leaves behind~$I=0$, indicating that the space 
is flat.

\paragraph*{Evolution of~$A=2.5$ centered Brill data:}
For larger, but still subcritical~$A$, the waves will 
interact nonlinearly before dispersing, but ultimately they 
still leave behind flat space. The Kretschmann scalar for centered
Brill data with~$A=2.5$ takes its maximum at the origin, but now with 
value of~$\approx 2320$. We evolved these data with~$\eta=2\approx 2/(5M)$.
The lapse at the origin again decreases at early times, this time to 
smaller values than for~$A=1$, and then returns to unity. The 
Kretschmann scalar also disperses to infinity, as before.

\begin{figure*}[t]
\centering
\includegraphics[]{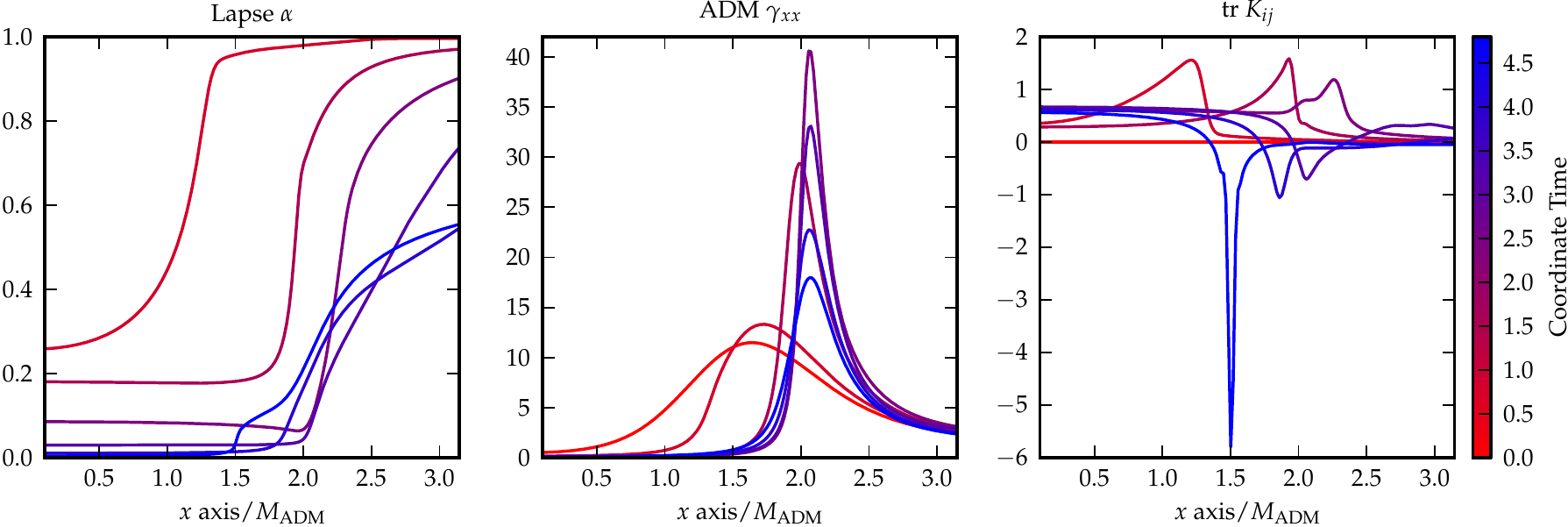}
\caption{Snapshots of the lapse function $\alpha$ (left panel), 
the ADM metric component $\gamma_{xx}$ (center panel) and
the trace of the extrinsic curvature $\mathrm{tr} K_{ij}$ at different
instants of time, for~$A = 5$.  We show all quantities along 
the~$x$ axis, with time indicated by hot-to-cold colors. The lapse 
develops a minimum on a ring of radius~$r \approx 2.0$, which 
then travels towards the origin. The spatial metric develops 
a large gradient around the same ring. A sharp feature at 
radius~$r \approx 1.5$ appears in the lapse, and~$K$ diverges 
around the same place. The latter feature is not visible in the 
metric.}\label{Fig:Brill_lapse_metric_profiles}
\end{figure*}

\paragraph*{Evolution of~$A=5$ centered Brill data:} For 
amplitudes larger than some critical~$A_\star$ one expects
black hole formation. For larger values of~$A$ our simulations 
are not successful, in the sense that we are not able 
to track the formation of an apparent horizon as it settles 
down to a Schwarzschild hole. We show more detailed results 
for~$A=5$ in Fig.~\ref{Fig:Brill_lapse_metric_profiles}, where 
profiles of the lapse function~$\alpha$, the metric 
component~$\gamma_{xx}$ and trace~$K$ are plotted at different 
instants of time. The lapse collapses in the central region and 
develops a minimum along a ring of radius~$r\approx 2.0$ in 
the equatorial plane in a simulation with~$\eta=11.4\approx 8/M$.
The metric simultaneously develops an increasingly large gradient 
across this ring, which ultimately 
turns into a discontinuity if we use~$\eta=0$ in the gamma-driver 
condition~\eqref{Gamma-driver}. Associated with this gradient is 
a large numerical error which can be seen, for example, in the 
violation of the constraints. A pulse in the lapse then approaches
the origin, so that the region with nearly-vanishing 
lapse becomes smaller. As this happens the trace of the extrinsic 
curvature at the incoming lapse pulse becomes large, negative 
and sharp, ending in a numerical failure that we believe is a 
coordinate singularity. Similar behavior has been observed with 
the~$1+\log$ gauge elsewhere, see for example~\cite{AlcMas97}.  
Furthermore, we were able to reproduce this failure in the spherical 
code used to develop the Z4c 
formulation~\cite{BerHil09,RuiHilBer10,WeyBerHil11,CaoHil11,HilBerThi12} 
by evolving flat space with a perturbed initial lapse with precisely 
the gauge of the Brill wave evolutions. This feature causes the 
numerical approximation to fail at around~$t=5.5$. Curiously here,
in preliminary tests, we found that mesh-refinement can cause problems. 
Often coarser grids are used to push the outer boundary far away 
inexpensively. But we found that the solutions being constructed 
are so extreme that if the grids are too coarse then they will fail 
during the single Runge-Kutta time-step needed before the data from 
the finer boxes can be used to overwrite coarse grid data. 

\paragraph*{Discussion, comparison with the literature:} The 
basic picture is similar to other nonlinear dynamical systems 
for which the strength of the initial fields is controlled by 
a parameter~$A$, which separates two distinct states at some 
critical value~$A_\star$.  Near this value such 
systems may display critical behavior, as reviewed 
elsewhere~\cite{Gun00,Gun02}. For vacuum spacetimes, critical 
collapse was first reported by~\cite{AbrEva92,AbrEva93}, who 
used Teukolsky data.  To the best of our knowledge,
all published results since then have adopted Brill initial data (see 
Table~\ref{Table:PreviousResults}.) 
Only one author, Sorkin~\cite{Sor10}, has reported critical phenomena for Brill data, 
and even those simulations required significant fine-tuning of 
parameters in the gauge conditions. The study also reports qualitative 
differences from the earlier work; in particular,~\cite{Sor10} 
finds that a spacetime singularity forms on a ring of non-zero 
radius in the equatorial plane, while for the simulations 
of~\cite{AbrEva92,AbrEva93} the singularity formed centrally.
Finally, Sorkin reports a critical value of~$A_\star\sim 6.27$, in 
contrast to smaller values found in earlier studies, for 
example~$A_\star\sim4.76$ in~\cite{San06}. The size of the 
difference is puzzling. Unfortunately, since we have not 
been able to evolve large data reliably with the moving-puncture 
gauge, we can not shed any light on the issue here. 

\paragraph*{Summary:} We conclude that moving-puncture coordinates 
are not suitable for the evolution of the Brill waves considered 
in this section. For several other coordinate choices, including 
maximal slicing for the lapse, as well as quasiisotropic or zero 
shift, it also appears to be difficult to obtain sufficiently reliable 
simulations that allow the study of critical phenomena in the 
vicinity of the critical amplitude. We know from other studies
that moving-puncture coordinates can be used in collapse 
scenarios, so rather than altering the gauge choice we will 
study evolutions of different initial gravitational wave data.

\subsection{Teukolsky Waves}
\label{subsection:Teukolsky}

\paragraph*{Initial data:} Linear solutions to Einstein's 
equations describing quadrupolar gravitational waves can be 
constructed from a seed function
\begin{align}
F(r,t) = F_1(t-r) + F_2(t+r),
\end{align}
where~$F_2$ describes an outgoing solution while~$F_1$ describes 
an ingoing solution (see~\cite{Teu82}; also Section~9.1 in~\cite{BauSha10}; 
see also~\cite{Rin08b} for a generalization to all multipoles).    
We choose~$F_1 = - F_2$ so that the resulting solution exhibits a 
moment of time symmetry, and hence~$K_{ij} = 0$, at $t=0$. We then 
choose
\begin{align} \label{Teukolsky_seed}
F_1(u) = \frac{A}{2} \left( \frac{u}{\sigma} e^{-((u + r_0)/\sigma)^2} +
         \frac{u}{\sigma} e^{-((u - r_0)/\sigma)^2} \right)
\end{align}
with~$u\equiv t-r$. Here~$A$ is again a measure of the wave's 
amplitude,~$\sigma$ of its wavelength, and~$\rho_0$ of the center 
of the original wave.  Following the prescription in~\cite{Teu82} we 
construct the spatial metric~$\gamma_{ij}$ for a 
quadrupolar~($\ell = 2$),~$m = 0$ wave from the seed 
function~\eqref{Teukolsky_seed}. The resulting metric satisfies 
Einstein's equations to linear order in~$A$.
In order to construct true initial data from the above solution 
we only need to solve the Hamiltonian constraint, since the 
momentum constraint~\eqref{mom} is satisfied identically by 
time-symmetric data. We do so by adopting the above metric 
at~$t=0$ as a conformally related metric~$\bar \gamma_{ij}$, and 
solving the Hamiltonian constraint~\eqref{Ham_conformal} for 
the conformal factor~$\psi$. Following~\cite{AbrEva92,AbrEva93} 
we consider ``non-central" initial data, i.e.~wave packets that 
are initially centered on a positive radius~$r_0$. Specifically, 
we choose~$r_0=2$ and~$\sigma = 1/2$ for all simulations presented 
in this section.

\paragraph*{Evolution of Teukolsky data:} For a given 
choice of~$\sigma$ and~$r_0$, the spacetime again depends 
only on the amplitude~$A$, although the specifics of the evolution 
will of course depend upon the gauge. Time-symmetric initial data 
represent a superposition of interacting ingoing and outgoing 
waves. At the amplitudes we consider, the interaction of these 
waves is weak. The outgoing wave travels toward infinity and does not 
play an important role in our simulations. The ingoing wave, on 
the other hand, travels toward the origin. For sufficiently small 
initial amplitudes the time evolution is well approximated by 
the analytical, linear solution. If the amplitude is smaller than 
a certain critical value~$A_{\star}$ we again expect that the evolution 
ultimately leaves behind flat space, possibly after interacting in 
a nonlinear fashion. For~$A > A_{\star}$ the waves collapse to form 
a black hole. Note that the wave's amplitude increases as it travels 
from~$r_0$ to the origin, so that we expect~$A_{\star}$ to be smaller 
for ``non-central" initial data than for similar ``central" initial 
data. We perform evolutions with moving-puncture coordinates. 
Our expectations for small initial data are borne out in 
practice, so we will not comment further.

\begin{figure}[t]
\centering
\includegraphics[]{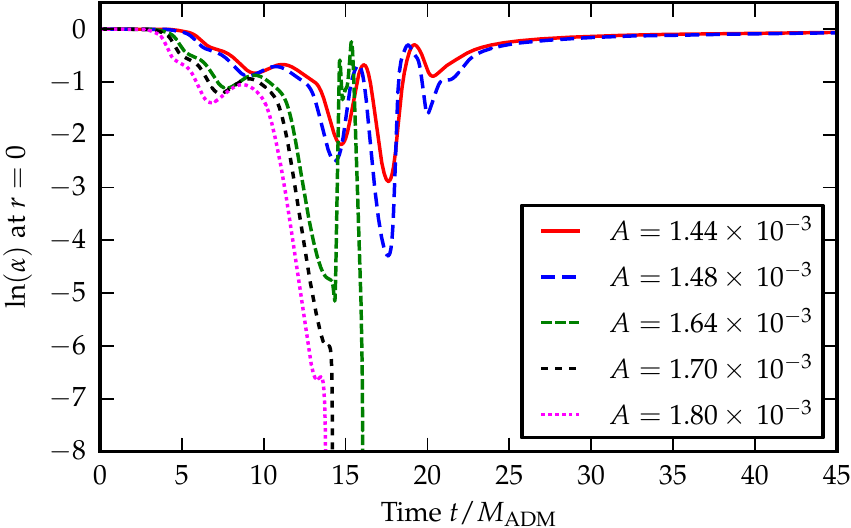}
\caption{The central value of the lapse as a function of time 
for Teukolsky waves with~$r_0 = 2$ and~$\sigma = 1/2$, evolved 
with moving-puncture gauge conditions.  We show results for 
different values of $A$.  For subcritical initial data 
with~$A < A_{\star} \approx 0.0015$, the wave disperses to spatial 
infinity and leaves behind flat space, while for~$A > A_{\star}$ 
the fields collapse to form a black hole.}
\label{Fig:Teukolsky_lapse_at_center}
\end{figure}

\begin{figure}[t]
\centering
\includegraphics[]{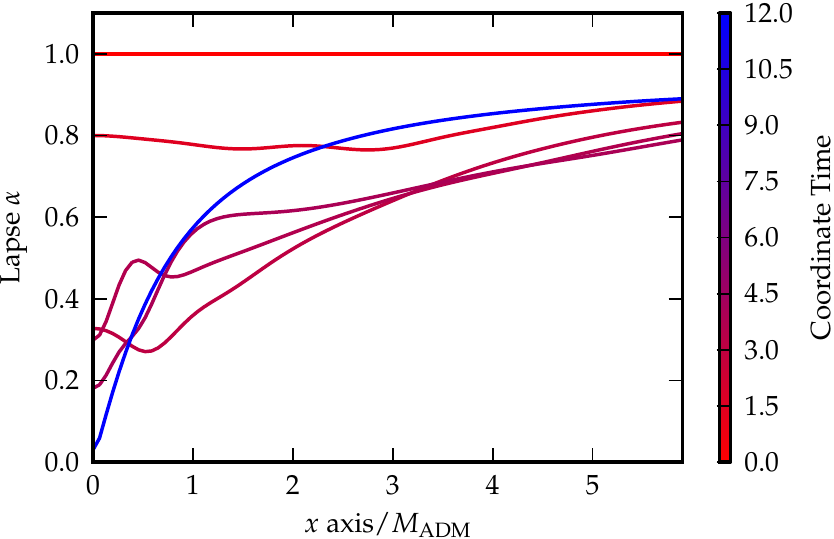}
\caption{Profiles of the lapse function $\alpha$ for a Teukolsky 
wave with amplitude~$A = 0.0018$.  Even though at early times 
the lapse forms a minimum at finite radius $r$, it ultimately 
collapses most rapidly at the center.  This behavior is qualitatively 
different from the behavior of the lapse in the collapse of a 
Brill wave, as shown in the top panel of 
Fig.~\ref{Fig:Brill_lapse_metric_profiles}.}
\label{Fig:Teukolsky_lapse_profile}
\end{figure}

\begin{figure}[t]
\centering
\includegraphics[width=0.35\textwidth]{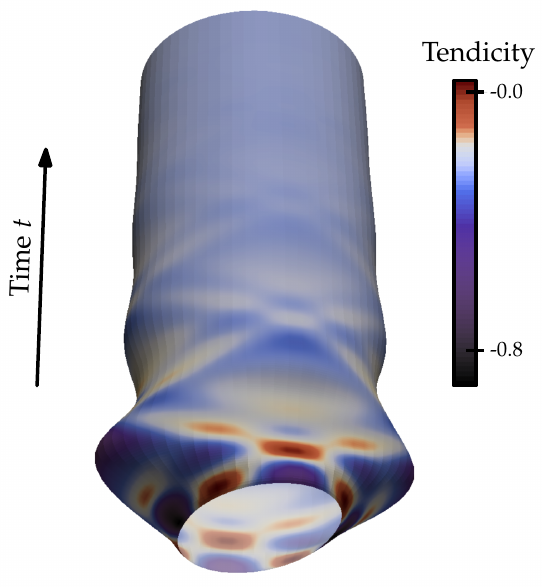}
\caption{A ``world tube" showing the newly formed horizon in 
the collapse of a Teukolsky wave for~$A=0.0018$. For each 
instant of time we show a ring that represents the coordinate 
location of the horizon along lines of constant longitude, 
going from one pole to the other and back.  Connecting these 
rings for different times results in the cylinder-like shape 
shown in the figures.  The shape of this cylinder demonstrates 
the horizon's initial quasi-normal oscillation, while the color 
coding shows the horizon's tendicity ${\mathcal E}_{NN} M^2$. 
After a few oscillations the horizon settles down into that of 
a static Schwarzschild black hole.}
\label{Fig:Teukolsky_horizon}
\end{figure}

\paragraph*{Numerical evolution of supercritical data:} The 
main result of this work is that we are able to evolve Teukolsky
wave initial data through black hole formation. As an example 
of such a supercritical evolution we focus on results 
for~$A=0.0018$. As shown in Fig.~\ref{Fig:Teukolsky_lapse_at_center}, 
the central value of the lapse exhibits several large-amplitude 
oscillations, but ultimately approaches zero at the center, 
indicating the formation of a black hole. In 
Fig.~\ref{Fig:Teukolsky_lapse_profile} we show profiles of the 
lapse at different instances of time. At early times, the lapse 
takes a minimum at finite radius, but at later times it 
collapses most rapidly at the center. This behavior should be 
compared with what we found for Brill waves, for which the lapse 
always collapses most rapidly at a finite radius (see the top 
panel in Fig.~\ref{Fig:Brill_lapse_metric_profiles}.) The 
peak of the curvature scalar occurs at the origin. We found 
that an apparent horizon forms at~$t\approx 4.2$. In 
Fig.~\ref{Fig:Teukolsky_horizon} we show the ``world tube" 
of this horizon in a spacetime diagram.

\begin{figure}[t]
\includegraphics[]{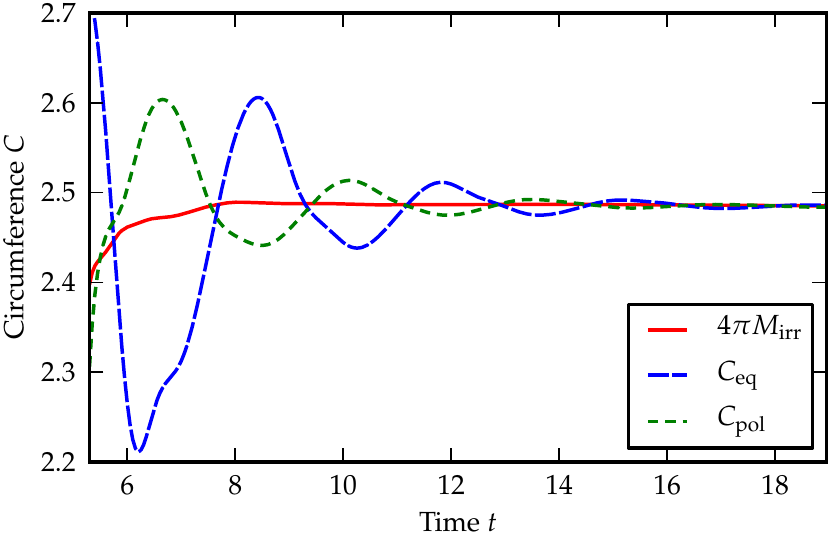}
\caption{The equatorial and polar proper circumference of the 
newly formed horizon of a Teukolsky wave for~$A=0.0018$. Also 
included is the irreducible mass $M_{\rm irr}$, multiplied 
by~$4\pi$, which equals the proper circumference of a spherically 
symmetric Schwarzschild black hole (see 
equation~\eqref{Schwarzschild_circum}.) The initially formed black 
hole is not spherical, but oscillates and settles down into a 
Schwarzschild black hole. During the initial oscillation the 
irreducible mass of the black hole still increases by a few 
percent.} \label{Fig:Teukolsky_hor_circum}
\end{figure}

\paragraph*{Collapse to a black hole and horizon formation:}
Our initial data are axisymmetric, but carry no angular momentum.  
Therefore, if a black hole forms in the time evolution of these 
initial data, this black hole must ultimately settle down into 
a Schwarzschild black hole.  However, since the initial data are 
not spherically symmetric, the newly formed black hole may also 
deviate from spherical symmetry.  We then expect that these 
deviations from spherical symmetry lead to quasi-normal 
oscillations in the horizon that damp away and leave behind 
a Schwarzschild black hole. This behavior can be observed 
in the horizon's world-tube in Fig.~\ref{Fig:Teukolsky_horizon}, 
which clearly shows these initial oscillations with 
a rapidly decreasing amplitude, leaving behind a spherical 
horizon at late times.  We note, however, that 
Fig.~\ref{Fig:Teukolsky_horizon} only shows the coordinate 
location of the horizon, which does not have an immediate physical 
meaning. As a (spatial) coordinate-independent measure of the 
horizon's deviation from spherical symmetry we plot its equatorial 
and polar proper circumferences in Fig.~\ref{Fig:Teukolsky_hor_circum}.  
Both circumferences perform a damped oscillation, almost exactly 
out of phase, and, as expected, settle down to the same, 
time-independent value.  The circumference of a Schwarzschild 
black hole is given by
\begin{align} \label{Schwarzschild_circum}
C = 2 \pi R = 4 \pi M,
\end{align}
where~$R = 2 M$ is the circumferential radius of a Schwarzschild 
black hole.  We therefore include the value~$4 \pi M_{\rm irr}$ in 
Fig.~\ref{Fig:Teukolsky_hor_circum}, where we compute the black 
hole's irreducible mass~$M_{\rm irr}$ from the proper area of its 
apparent horizon. Fig.~\ref{Fig:Teukolsky_hor_circum} shows that 
both the polar and equatorial proper circumferences settle down 
to the Schwarzschild circumference, demonstrating that the 
initially distorted black hole settles down to the trumpet geometry. 
The fact that the late-time horizon takes a spherical shape even in 
coordinate space, as seen in Fig.~\ref{Fig:Teukolsky_horizon}, 
demonstrates that the Gamma-driver shift condition~\eqref{Gamma-driver} 
allows the coordinate evolution to reflect the spherical symmetry 
of the spacetime.  

\paragraph*{Horizon tendicity:} As an alternative (spatial) 
coordinate-independent measure of the horizon geometry we 
compute its tendicity,
\begin{align} \label{tendicity}
{\mathcal E}_{NN} \equiv {\mathcal E}_{ij} s^i s^j,
\end{align}
where~${\mathcal E}_{ij}$ is the electric part of the Weyl tensor, 
and $s^i$ the spatial unit normal on the horizon 
(see \cite{OweBriChe11,ZhaZimNic12,NicZimChe12}; see 
also~\cite{DenBau12} for an analytical demonstration.)  For a 
Schwarzschild black hole the tendicity is
\begin{align} \label{tendicity_Schwarzschild}
{\mathcal E}_{NN}^{\rm SS} = - \frac{1}{4 M^2}
\end{align} 
(see, e.g., \cite{OweBriChe11,DenBau12}). In 
Fig.~\ref{Fig:Teukolsky_horizon} we have indicated the horizon 
tendicity using a color coding.  At early times, the tendicity 
varies significantly across the horizon, meaning that the horizon 
is significantly distorted, while at late times the tendicity 
becomes increasingly uniform across the horizon, with a value 
close to the analytical value~\eqref{tendicity_Schwarzschild} for 
a Schwarzschild black hole.   As the newly formed black hole 
performs quasi-normal oscillations, deviations of the tendicity 
from its average value propagate across the horizon in a wave-like 
manner, from the poles to the equator and back. This can be seen 
in the color-coding in Fig.~\ref{Fig:Teukolsky_horizon}. For
for a detailed analysis of quasi-normal modes using the 
tendex-vortex formalism see~\cite{NicZimChe12} .

\begin{figure}[t]
\begin{center}
\includegraphics[]{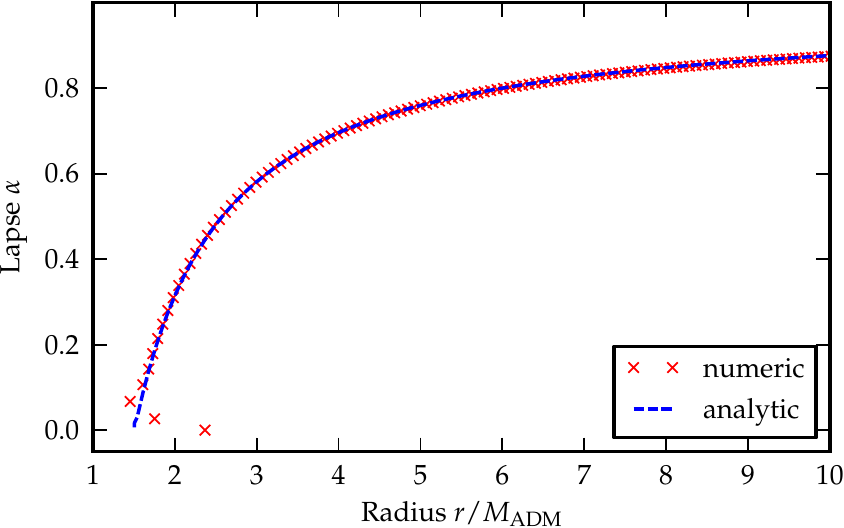}
\end{center}
\caption{Radial profile of the lapse~$\alpha$ for the collapse
of a Teukolsky wave with~$A = 0.0018$, at time~$t=15.8$ (crosses). 
This evolution is carried out with ``non-advective" 1+log slicing, 
so that we can easily compare with the analytical solution for 
a maximally sliced trumpet solution (solid line).  Note that 
the overall scale of the analytical solution for the lapse is 
set by the boundary condition at~$r=\infty$. The numerical 
solution, however, results from the collapse of non-linear wave 
initial data, and is not affected by the boundary condition 
at spatial infinity; accordingly, the overall factor of the 
solution for the lapse may be different. Here we found excellent 
agreement by multiplying the analytical solution with a factor 
of 0.98. The inner-most few grid-points are affected by numerical 
noise that results from finite-differencing across the singularity 
at~$r = 0$.}\label{Fig:lapse}
\end{figure}

\paragraph*{Approach to a trumpet slice:} To complete this 
section we analyze the late-time solution to which our 
dynamical evolutions settle down. Given that we use 
moving-puncture coordinates, we expect that a black hole settles 
down into a trumpet 
geometry~\cite{HanHusPol06,HanHusOMu06,BauNac07,HanHusOhm08}. 
In order to compare results with the analytical solutions for a 
maximally sliced trumpet solution~\cite{BauNac07}, we show in 
Figs.~\ref{Fig:lapse} results for an evolution with the 
``non-advective" version of the 1+log slicing~\eqref{1+log} 
and Gamma-driver~\eqref{Gamma-driver} (we also used $\mu_S = 3/4$ 
and $\eta = 0$ in (\ref{Gamma-driver}) for these simulations). 
In Fig.~\ref{Fig:lapse} we show a snap-shot of the lapse at 
time~$t=15.8$, and compare our numerical results with the 
analytical results of \cite{BauNac07}.   We plot the lapse as 
a function of areal radius $R$ in order to compare gauge-invariant 
quantities.  The agreement is excellent, after an over-all scale 
factor has been adjusted.  This scale factor allows for the 
fact that the analytical solution assumes that the lapse 
approaches unity at spatial infinity, while our dynamical 
simulations do not impose this condition. We compare numerical 
and analytical values for the shift, and find similarly good 
agreement.  

\paragraph*{Experiments near the critical region:} As we increase 
the amplitude from~$0$, we observe increasingly large 
and rapid oscillations. This can be seen in 
Fig.~\ref{Fig:Teukolsky_lapse_at_center}, where we show central 
values of the lapse as a function of time for different values 
of the amplitude~$A$. For the simulations shown in 
Fig.~\ref{Fig:Teukolsky_lapse_at_center} we adopted the 
moving-puncture coordinates~\eqref{1+log} and~\eqref{Gamma-driver} 
with~$\eta \approx 2/M_{\rm ADM}$. We experimented with the 
non-advective version of the 1+log slicing condition~\eqref{1+log} 
and found similar results. We summarize our results in 
Table~\ref{Table:CriticalPoint_Teukwave}, where we list the 
initial ADM mass, and the irreducible mass of the black hole 
if such a black hole forms. As can be seen in 
Table~\ref{Table:CriticalPoint_Teukwave} and 
Fig.~\ref{Fig:Teukolsky_lapse_at_center}, the critical value 
of the amplitude is between~$A=0.00148$ and~$A=0.00164$. 
For~$A > A_{\star}$ the lapse ultimately collapses to zero at 
the center. For both sub-- and supercritical values of the 
amplitude, the lapse exhibits more and larger oscillations 
before returning to unity or collapsing to zero for amplitudes 
closer to the critical value. This behavior is an 
indication of the critical behavior in the vicinity of the 
critical amplitude.

\begin{table}[t]
  \begin{ruledtabular}
    \begin{tabular}{c|c|c|c}
      \hline
$A$ & $M_{\rm irr}$ & $M_{\rm ADM}$ & $M_{\rm irr}/M_{\rm ADM}$  \\
\hline
0.00140 & --     & 0.222 & --     \\
0.00142 & --     & 0.229 & --     \\
0.00144 & --     & 0.236 & --     \\
0.00146 & --     & 0.243 & --     \\
0.00148 & --     & 0.250 & --     \\
0.00164 & 0.144 & 0.312 & 0.460 \\
0.00166 & 0.151 & 0.321 & 0.470 \\
0.00168 & 0.158 & 0.329 & 0.480 \\
0.00170 & 0.164 & 0.338 & 0.513 \\
0.00175 & 0.181 & 0.360 & 0.550 \\
0.00180 & 0.197 & 0.383 & 0.584 \\
0.00200 & 0.265 & 0.486 & 0.737 \\
\hline
\end{tabular}
  \end{ruledtabular}
\caption{Summary of results for non-linear Teukolsky waves with 
different initial amplitude $A$. The amplitudes~$A=0.00148$, 
and~$A=0.00164$ are respectively the highest subcritical and 
lowest supercritical data we tried which did not fail. We tabulate 
the irreducible mass~$M_{\rm irr}$ of the black hole, if a black hole 
formed in the evolution, the initial ADM mass~$M_{\rm ADM}$, and 
the ratio between the two. The irreducible mass increases while the 
initially formed black hole settles down into equilibrium 
(compare Fig.~\ref{Fig:Teukolsky_hor_circum}); we list here 
the near-equilibrium value after several oscillation periods.}
\label{Table:CriticalPoint_Teukwave}
\end{table}

\paragraph*{Summary:} We conclude 
that, at least for amplitudes not too close to the critical 
amplitude, moving-puncture coordinates are well-suited to 
simulate the evolution of Teukolsky waves.  For subcritical 
waves we follow the nonlinear interaction of the waves and their 
dispersal to spatial infinity, while for supercritical waves 
we track their implosion, detect the formation of apparent horizons, 
and follow the evolution of the newly-formed black holes as they 
settles down to spherical symmetry. For amplitudes close to 
the critical amplitude our codes fail. We believe that this is 
caused by the lack of sufficient resolution as, close to the 
critical amplitude, the dynamical evolution leads to increasingly 
small features. 

\section{Characterization of Initial Data}
\label{section:Character}

Evolving Brill and Teukolsky waves in Sec.~\ref{section:Waves} 
we found that Teukolsky data can be evolved with moving-puncture 
coordinates without problems -- except, possibly, in the immediate 
proximity of the critical point -- whereas Brill data are more 
troublesome. While it is difficult to pin-point what exactly 
creates this difference in behavior, we offer some speculations 
on the causes of these differences in this Section. An obvious 
set of questions present themselves. The Brill data 
were centered at the origin, whereas the Teukolsky data were not.
Could this be the cause of the difference? Are the two types 
of data somehow geometrically different? If so, would this 
difference be maintained in time evolution, and can we modify the 
Brill data to make it more amenable to numerical evolution with
the moving-puncture gauge? We address each of these issues in 
turn.

\subsection{Off-center Brill wave evolutions}
\label{subsection:Off-Brill}

\paragraph*{Small data:} We evolve two sets of weak, off-center Brill wave initial data 
with off-set~$\rho_0=4$ in~\eqref{Brill_seed}.  Specifically, we choose data with
amplitudes~$A=0.053$ and $A=0.0815$, which have ADM 
masses~$\approx0.59$ and~$\approx 1.4$ respectively. For the 
weaker~$A=0.053$ data we use~$\eta\approx1/M$, while for~$A=0.0815$ we 
chose~$\eta\approx2/M$.  As for earlier weak Brill data we found that 
the both waves disperse after a brief interaction around the origin. 
For the weaker data, the Kretschmann scalar takes a 
maximum value~$\approx48$ at the origin, while for the stronger data the 
maximum is~$\approx200$, again at the origin. 

\begin{figure}[t]
\centering
\includegraphics[]{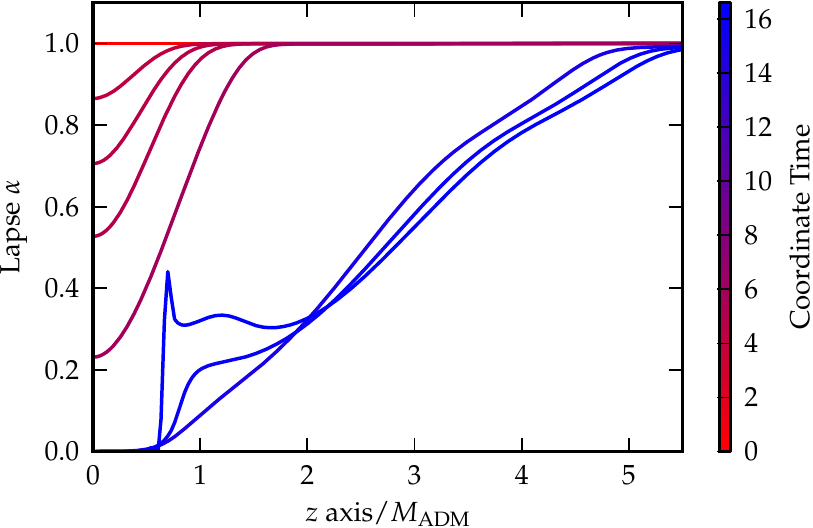}
\caption{Profiles of the lapse at different instances of time along the $z$ axis
for a off-centered Brill wave with $A=0.12$ and $\rho_0=4.0$. Initially the
lapse collapses at the center. At late times of the evolution, a gauge pulse
is traveling in and gets blue-shifted. At the interface between the collapsed
lapse and the gauge pulse, a coordinate singularity appears.
This leads to a failure of the simulation.}
\label{Fig:coord_sing}
\end{figure}

\paragraph*{Large data:} We now take the same off-set 
but~$A=0.12$, which makes the ADM mass~$\approx3.15$. We used~$\eta=8/M$.
As in the ``small data'' tests, the lapse initially 
decreases most rapidly around the peak of the seed function in 
the~$xy$-plane. It then decreases to zero near the origin. 
Then an incoming gauge wave travels along the~$z$-axis towards 
the origin. Since the speed of the gauge wave is~$\sim\sqrt{2\alpha}$,
and travels from a region with~$\alpha\sim1$ to one with~$\alpha\sim0$,
there is a rapid blueshift effect; the solution becomes badly 
resolved. At some point in time, around the interface of 
these two regions, the trace of the extrinsic curvature becomes 
negative, which leads to an increase in the lapse (see eq.~\eqref{1+log}.)
Ultimately, this results in a coordinate singularity
causing the code to fail at~$t\approx16$.  We note a remarkable similarity
between the lapse profile we obtain, shown in Fig.~\ref{Fig:coord_sing}, and 
that shown in Fig.~$2$ of~\cite{AlcMas97} where such coordinate 
singularities were studied in evolutions of flat-space. We conclude 
that, for sufficiently large amplitudes, moving-puncture coordinates 
fail even for off-centered Brill wave initial data.

\subsection{Axisymmetric twist-free, time symmetric data}
\label{subsection:Character_data}

\paragraph*{Harmonic spatial coordinates:} We start by searching 
for differences in the geometry of Brill and Teukolsky 
waves. In cylindrical coordinates, the conformally related 
metric for Brill wave initial data is given by
\begin{align} \label{Brill_conf}
\bar \gamma_{ij} = 
\left(
\begin{array}{ccc}
e^{q}&0&0\\
0&e^{q}&0\\
0&0&\rho^2
\end{array}\right)
\end{align}
(see eq.~(\ref{Brill_metric})), whereas for $m=0$ Teukolsky 
waves the conformally related metric takes the form
\begin{align} \label{Teu_conf}
\bar \gamma_{ij} = 
\left(
\begin{array}{ccc}
\bar{\gamma}_{\rho\rho}&\bar{\gamma}_{\rho z}&0\\
\bar{\gamma}_{\rho z}&\bar{\gamma}_{zz}&0\\
0&0&\bar\gamma_{\phi\phi}
\end{array}\right)\,.
\end{align}
Evidently, the two data sets are 
given in different coordinate systems, and a meaningful comparison 
can only be made once they have been expressed in the same 
coordinates. However, any axisymmetric, twist-free metric can be 
brought into the form
\begin{align} \label{axi_metric}
\bar \gamma_{ij} = 
\left(
\begin{array}{ccc}
e^q & 0 & 0 \\
0 & e^q & 0 \\
0 & 0 & \rho^2 V 
\end{array}\right)\,,
\end{align}
in some coordinate system~$(\varrho,\xi,\phi)$. Crudely speaking, 
this is possible because the two-metric in the~$\rho$-$z$ 
subspace can always be brought into an explicitly conformally 
flat form~\cite[Ch.~3, Ex.~2]{Wal84}. 

\paragraph*{Geometrically oblate and prolate initial data:} Any 
spherically symmetric metric can be brought into the 
form~\eqref{axi_metric} with~$V=e^q$. For our gravitational wave 
initial data, which are not spherically symmetric,~$V$ will, in 
general, be different from~$e^q$. Evidently, deviations of~$V$ 
from~$e^q$ can be 
produced in two ways: either $V > e^q$ or $V < e^q$. We 
characterize data with~$V > e^q$ as {\it geometrically oblate} 
and data with~$V< e^q$ as {\it geometrically prolate}. We use 
the word `geometrically' to distinguish the terminology from 
that normally used with Brill waves, where the word oblate or 
prolate applies to the seed function. Clearly, both~$V$ 
and~$e^q$ are functions of the coordinates, so that data may 
be geometrically oblate in some region and prolate in another. 
We also point out that we apply this characterization only to 
the initial data; it is not evident whether or how this 
characterization is maintained during a time evolution, even 
if the data are globally geometrically oblate or prolate 
initially. The characterization as geometrically oblate or 
prolate may nevertheless be a useful distinction between the 
geometries of Brill and Teukolsky data evolved earlier. An example 
of initial data of a certain character which is maintained is 
given in~\cite{Chr08}. 

\paragraph*{Geometric oblateness of Brill and Teukolsky data:} 
For the Brill wave initial data of Section~\ref{subsection:Brill} 
we have~$V=1$; moreover we chose a positive amplitude~$A$ in the 
seed function~\eqref{Brill_seed}, which results in~$e^q \geq 1$. 
This means that the Brill wave initial data of 
Section~\ref{subsection:Brill} are geometrically 
prolate everywhere except on the~$z$-axis. For Teukolsky waves the 
classification is less obvious, because it requires an additional 
coordinate transformation. Interestingly, however, Fig.~3 
in~\cite{AbrEva93} shows that the geometry close to the center 
is geometrically oblate. (The quantity~$e^\eta$ used by~\cite{AbrEva93} 
is a measure of the ratio~$e^q/V$; the fact that their~$\eta$ is 
negative in a region around the origin implies that the geometry 
is oblate there.)  This observation could point to a fundamental 
difference in the geometries of~$A>0$~Brill and Teukolsky 
waves.

\paragraph*{Discussion:} These arguments are neither rigorous 
nor complete, but they lead to an immediate suggestion: if it 
were true that geometrically oblate data are better behaved in dynamical 
evolutions than prolate data -- for example in the sense that 
they form a singularity at the center rather than on a ring -- 
then it would be of interest to produce geometrically oblate 
Brill wave initial data. We consider such data next.

\subsection{Negative-amplitude Brill waves}
\label{subsection:NegativeAmpBrill}

\begin{figure}[t]
\centering
\includegraphics[]{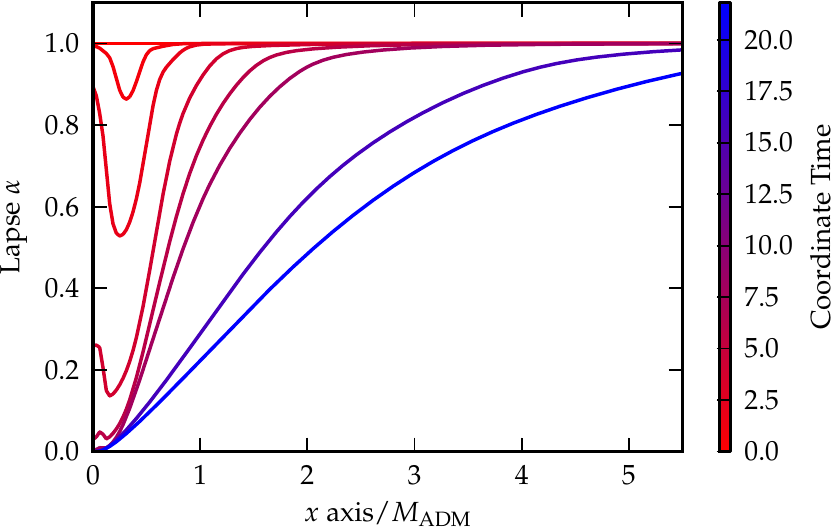}
\caption{Profiles of the lapse at different instances of time for 
a negative-amplitude, centered Brill wave with $A = -5$.  These profiles
should be compared with (a) those of a positive-amplitude Brill 
wave in the top panel in Fig.~\ref{Fig:Brill_lapse_metric_profiles}, 
and (b) those for a Teukolsky wave in Fig.~\ref{Fig:Teukolsky_lapse_profile}.}
\label{Fig:Brill_neg_amp_lapse_profile}
\end{figure}

\paragraph*{Geometrically oblate Brill waves:} In this section 
we consider geometrically oblate Brill wave data. Such data can 
be produced in exactly the same way as the prolate data in 
Section~\ref{subsection:Brill}, by adopting negative 
amplitudes~$A < 0$ in the seed function~\eqref{Brill_seed}.  

\paragraph*{Centered geometrically oblate Brill waves:} We 
evolved three sets of initial data with~$\rho_0=0$ and $A<0$, 
namely~$A=-1,\,-2.5$ and~$A=-5$, with ADM masses of 
approximately~$0.61,1.37$ and~$3.15$ respectively. As in the 
geometrically prolate case, the first two data sets just allow 
the Kretschmann scalar to propagate away, taking maximal values
of about~$56$ and~$1072$ appearing at the origin. For the 
stronger field evolution the numerics again eventually fail, 
although now at around~$t=22.5$. In 
Fig.~\ref{Fig:Brill_neg_amp_lapse_profile} we show profiles 
of the lapse at different instances of time for such a  
Brill wave with~$A = -5$. Interestingly, these profiles 
are qualitatively different from those for the geometrically 
prolate, positive-amplitude Brill waves shown in the top panel 
of Fig.~\ref{Fig:Brill_lapse_metric_profiles}. While for 
geometrically prolate Brill waves the lapse always takes a 
minimum at finite radius, which we found to coincide with the 
development of increasingly large gradients in the spatial 
metric, for negative-amplitude Brill waves the lapse ultimately 
takes a minimum at the center. This is the same behavior that 
we highlighted for Teukolsky waves in 
Fig.~\ref{Fig:Teukolsky_lapse_profile}. We were able to 
follow the collapse of geometrically oblate Brill wave past 
black hole formation, and in contrast to positive-amplitude 
Brill waves we were also able to locate apparent horizons, at 
least in the spherical coordinate code~\cite{BauMonCor12}. 
However, at least with the setup we chose, the BAM apparent horizon 
finder did not give reliable results. At late times, steep gradients 
in the metric functions again appeared, but this time well inside 
the horizon, close to the center. Ultimately, these gradients spoil 
further numerical evolution.

\begin{figure}[t]
\centering
\includegraphics[]{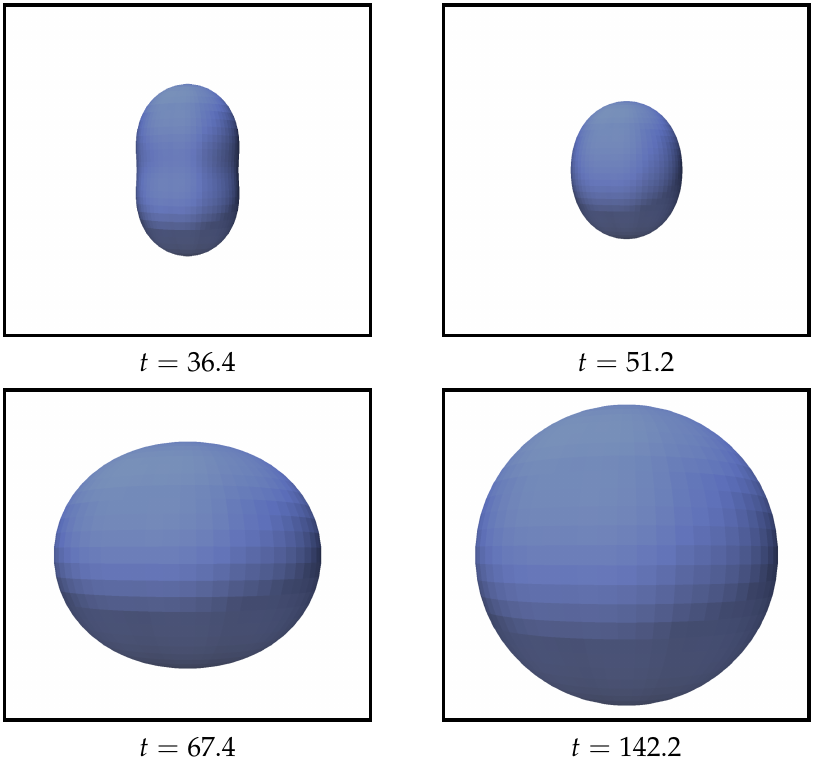}
\caption{Snap-shots of the apparent horizon for supercritical 
off-center geometrically oblate Brill waves. The apparent 
horizon is first discovered at~$t=36.4$ and has a peanut like 
shape, where in the plots the $z$-axis runs vertically. 
The subsequent frames show the oscillations of the horizon, 
which jumps at~$t\approx 65$.}\label{Fig:Brill_neg_amp_AH}
\end{figure}

\paragraph*{Off-center geometrically oblate Brill waves:}
We again evolved three sets of initial data~$A=-0.044,-0.061$ 
and~$A=-0.08125$, all with~$\rho_0=4$. The ADM masses of these 
spacetimes are approximately~$0.61,1.36$ and~$3.15$ with peak 
values of the Kretschmann scalar initially around~$64$, $108$, 
and~$88$ respectively. It is interesting that the ``larger''
initial data does not have the largest initial Kretschmann, but 
the evolution leaves no doubt that the~$A=-0.08125$ data is 
indeed the stronger. As in our previous results the two weaker 
data sets leave behind~$I=0$, with greater oscillations in the 
Kretschmann scalar in the~$A=-0.061$ case. The maximum of the 
Kretschmann scalar in the evolution of the~$A=-0.044$ is around~$19$ 
and occurs at the origin at~$t\simeq7.4$. Likewise with 
the~$A=-0.061$ data, the maximum occurs at the origin, with a 
value around~$373$ at~$t\simeq12$. In both the~$A=-0.044$ 
and the~$A=-0.061$ evolutions the Kretschmann scalar propagates 
away predominantly along the symmetry axis. We 
evolved the strongest data~$A=-0.08125$ set in the BAM code 
until~$t=150$. It collapses to form a black hole similarly to the 
Teukolsky data presented in section~\ref{subsection:Teukolsky}. 
An apparent horizon was first discovered at~$t=36.4$. In 
Fig.~\ref{Fig:Brill_neg_amp_AH} some snapshots of the evolution of 
the horizon are plotted. The apparent horizon mass eventually settles 
down to~$M=1.73$. Comparing the maximum resolution in this simulation 
relative to this scale with the earlier Teukolsky wave BAM evolution 
for $A=0.00175$, the present data have roughly nine times the resolution, 
which may explain why we did not obtain reliable results from the BAM 
apparent horizon finder earlier. So by 
choosing the parameters in the Brill wave data carefully we can obtain 
evolutions comparable to those we had with Teukolsky initial 
data.

\section{Summary}\label{section:Summary}

We presented numerical simulations of nonlinear gravitational 
waves. We adopted two different types of initial data -- 
Brill and Teukolsky waves -- and evolved them with two 
independent numerical codes.  

We consistently find that positive amplitude Brill waves,
most commonly evolved in the literature, fail to produce 
long-term stable evolutions with the moving-puncture gauge,
unless the initial amplitude is small. Evolving these data with 
this gauge leads to steep gradients in metric functions, which 
ultimately spoil the numerical evolution. Comparing with earlier
studies it seems most likely that the failure is a coordinate 
singularity. For positive-amplitude Brill waves we are also unable 
to locate black hole horizons, even for data that we believe do 
form black holes.

On the other hand, we find that Teukolsky waves do allow a 
stable, long-term evolution in moving-puncture coordinates, 
unless the initial amplitude is close to critical. We followed 
the evolution of the waves and their collapse to a black hole, 
tracking the newly formed horizon, and confirming that the 
spatial slices settle down to a trumpet geometry.

The primary motivation was therefore to provide one more example of 
successful simulations with moving-puncture coordinates that track 
the collapse of regular initial data to a black hole. Another 
motivation is to point out the surprising, qualitative differences
between Brill and Teukolsky wave initial data. We speculate that 
the choice of initial data has significantly contributed to the fact 
that the original studies of criticality in the collapse of nonlinear 
waves~\cite{AbrEva93} have been so difficult to reproduce.  
In retrospect, it is surprising that none of the studies after~\cite{AbrEva93} 
considered Teukolsky waves. Apart from choosing Brill waves for technical 
convenience, this may appear justified because critical phenomena are 
not expected to depend on details of the initial data, which is an aspect 
of universality found in many studies. However, wave collapse in 
axisymmetry also marks a departure from spherical symmetry often used in 
other studies. Axisymmetry opens up the possibility that new geometric 
aspects matter, and the present study is an example. 

Our findings also raise new questions. In particular, it would be desirable 
to understand why Brill and Teukolsky waves behave so differently. We 
discussed a characterization of twist-free, axisymmetric data as either 
geometrically prolate or oblate. While this characterization does point to 
qualitative differences in the respective geometries, our analysis is 
incomplete in the sense that it does not consider the time-dependence of 
the characterization. We believe that it would be well-worth to further 
pursue this or similar approaches in order to gain a deeper insight into the 
geometries of these waves. On the basis of this characterization, as an attempt 
to make the Brill wave data as close as possible to the Teukolsky evolutions,
we evolved off-center negative amplitude Brill waves. We found that we were
once again able to evolve such data through apparent horizon formation, and 
furthermore as they settle down to a Schwarzschild black hole.

Unfortunately, our simulations still break down, even for Teukolsky waves, in 
the most interesting regime, namely close to the critical amplitude.  This prevents 
us from analyzing critical phenomena with our current simulations. We believe 
that this failure occurs because of the lack of sufficient spatial grid resolution. 
As one approaches the critical point, the evolution leads to oscillations on 
increasingly small scales. At least with moving-puncture coordinates, the 
under-resolution of these oscillations causes the lapse to become negative, which 
then spoils the numerical evolution.  The need for increasingly fine spatial 
resolution in the vicinity of the critical point is not a new revelation, of course; 
it explains, for example, why a well-adjusted adaptive grid-refinement proved so 
crucial in the original simulations of Choptuik \cite{Cho93}. We also cannot exclude 
the possibility that the moving-puncture coordinates themselves fail close to the 
critical point, irrespectively of resolution. In either case, we plan to develop 
techniques (for example unequal grid-spacing in our spherical-coordinate code), and 
experiment with the slicing and gauge conditions, in order to study critical 
phenomena in the collapse of nonlinear waves in the future.

\acknowledgments

We are grateful to David Garfinkle for very interesting 
discussions. TWB gratefully acknowledges support from the 
Alexander-von-Humboldt Foundation, and would like to thank the 
Max-Planck-Institut f\"ur Astrophysik for its hospitality.  
This work was supported in part by the Deutsche 
Forschungsgemeinschaft (DFG) through its Transregional Center 
SFB/TR7 ``Gravitational Wave Astronomy'', by NSF grant 
PHY-1063240 to Bowdoin College, by the DFG Research Training 
Group 1523/1 ``Quantum and Gravitational Fields'', the Graduierten-Akademie Jena,  
and by the Spanish Ministry of Science through grant AYA2010-21097-C03-01. 
Computations were performed primarily at the LRZ (Munich). 

\begin{appendix}

\section{Numerics}
\label{section:Numerics}

\begin{figure}[t]
\centering
\includegraphics[]{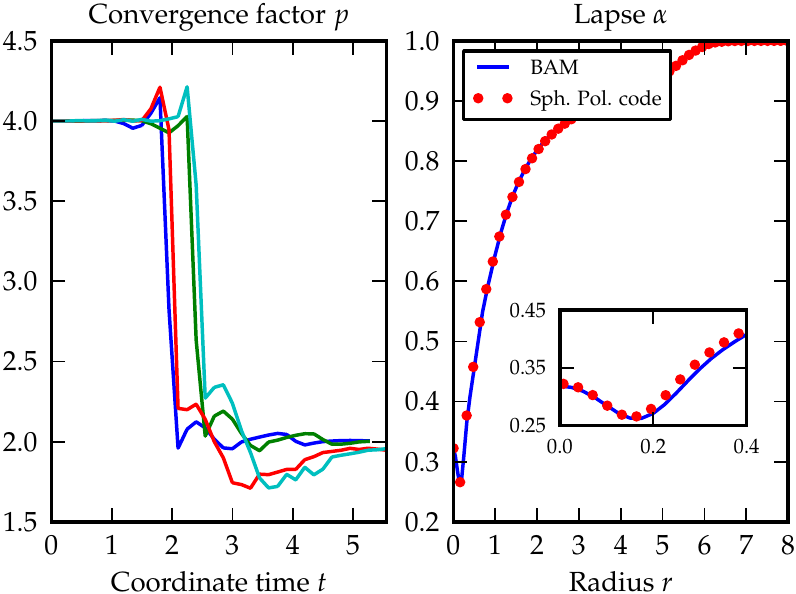}
\caption{In the left panel self-convergence tests for 
the~$\chi=\psi^{-4}$ variable are plotted from the BAM data. A 
centered Brill wave with~$A=5$ is evolved using four different 
resolutions. The red lines shows the self-convergence factor 
comparing~$170$,~$227$ and~$340$ spatial points along the axes 
in one of the mesh-refinement boxes with $x_{max}=8.7$. For the 
blue line we use~$227$,~$340$ and~$680$ points along the axes. 
Initially the simulation shows~$4$th order convergence. Note 
that the Berger-Oliger time interpolation at the mesh refinement 
boundary is done at second order, and may become the the dominant 
error when evolving strongly dynamic waves. Evidence for this is 
given by the green and turquoise lines, in which the convergence 
test is performed in a slightly smaller part of the grid. In the 
right panel a comparison between the numerical results obtained 
with our two codes, for a Teukolsky wave with amplitude~$A=0.0018$ 
is plotted. We show the lapse function~$\alpha$ at time~$t\approx 3.0$ 
as a function of~$r$ in the equatorial plane. The inset shows 
detail around the turning point.}\label{Fig:conv_comp}
\end{figure}

The numerical results presented in this paper were produced 
with two different numerical codes.  Both of these codes 
evolve Einstein's equations in the BSSN 
formulation~\cite{NakOohKoj87,ShiNak95,BauSha98} using 
finite-difference methods, but in completely independent 
implementations.

\paragraph*{The BAM code:} One of our codes is the BAM code, 
which adopts Cartesian coordinates and is described 
in~\cite{BruGonHan06,BruTicJan03,ThiBerBru11,HilBerThi12}.
The evolutions in this paper were performed with an explicit
4th-order Runge-Kutta method and 4th-order finite differences
for the spatial derivatives. Mesh refinement is provided 
by a hierarchy of cell-centered nested Cartesian grids and 
Berger-Oliger time stepping. Metric variables are interpolated 
in space by means of 6th-order Lagrangian polynomials. 
Interpolation in Berger-Oliger time stepping is performed at 
2nd order. In Fig.~\ref{Fig:conv_comp} we show a convergence
test for a Brill wave with positive amplitude $A=5$.

\paragraph*{The spherical-polar coordinate code:} Our other code is 
an implementation of BSSN in spherical polar 
coordinates~\cite{BauMonCor12}.  The code adopts a 
reference-metric formulation of the BSSN formalism~\cite{Bro09}, 
uses a partially-implicit Runge-Kutta (PIRK) method for the 
time evolution~\cite{MonCor12,CorCer12}, and scales out 
appropriate factors of~$r$ and~$\sin\theta$ from all tensorial 
quantities.  Spatial derivatives are evaluated using 4th-order 
finite differencing, except for advective (shift) terms, which 
are evaluated to 3rd order.  Our current implementation differs 
from that described in~\cite{BauMonCor12} in that it now uses a 
3rd order finite-differencing for the advective terms (rather 
than 2nd order), in that we have implemented an apparent horizon 
finder using the approach of~\cite{Shi97,ShiUry00}, and that we 
now use trilinos software~\cite{url:trilinos,HerBarHow05} to 
solve elliptic equations.  While this code does not make any 
symmetry assumptions, the axisymmetric solutions considered in 
this paper can be computed efficiently by choosing the minimum 
number of grid points possible in the $\phi$-direction.

\paragraph*{Comparison:} The codes produce consistent results. 
As an example, we compare in Fig.~\ref{Fig:conv_comp} the
lapse function~$\alpha$ in the equatorial plane at a 
time~$t\approx3$ for a collapsing Teukolsky wave.  

\end{appendix}

\bibliographystyle{unsrt}


\end{document}